\documentclass[pre,twocolumn,showpacs,preprintnumbers,amsmath,amssymb]{revtex4-1} 

\usepackage{dcolumn}
\usepackage{bm}
\usepackage[dvipsnames]{xcolor}
\usepackage{amsmath}
\usepackage{amssymb}
\usepackage{ifsym}
\usepackage{amsfonts} 
\usepackage{wasysym}  
\usepackage{graphics}  
\usepackage{psfrag} 
\usepackage{graphicx} 
\usepackage{epsfig}    
\usepackage{rotating}  
 \usepackage[latin1]{inputenc}
 \usepackage{color}
 \usepackage{rotating}
 \usepackage{multirow}
\usepackage{csquotes}
\usepackage{hyperref}
\usepackage[FIGTOPCAP]{subfigure}
\usepackage{stackengine}

\usepackage{color}
\usepackage{epsfig}
\definecolor{darkgreen}{rgb}{0,0.5,0} 
\definecolor{violet}{rgb}{0.5,0,0.5}
\definecolor{orange}{rgb}{0.2,0.5,0.5}
\newcommand{\avg}[1]{\langle #1\rangle}

\newcommand{\tn}[1]{\textnormal{#1}}
\newcommand{\labfig}[1]{\label{fig:#1}}

\newcommand{\reffig}[1]{\ref{fig:#1}}

\newcommand{\bequ}{\begin{equation}}
\newcommand{\eequ}{\end{equation}}
\newcommand{\bequa}{\begin{eqnarray}}
\newcommand{\eequa}{\end{eqnarray}}
\newcommand{\bse}{\begin{subequations}}
\newcommand{\ese}{\end{subequations}}

\usepackage{mathtools}

\usepackage{wrapfig}

\begin{document}

\title{Traffic Dynamic Instability}

\author{Louis Reese$^{1,2}$}\email{l.reese@tudelft.nl\\Present address: Department of Bionanoscience, Kavli Institute of Nanoscience, Delft University of Technology, Delft, The Netherlands} 
\author{Anna Melbinger$^{1,3}$}
\author{Erwin Frey$^{1,2}$}
\email{frey@lmu.de}
\affiliation{
$^1$Arnold-Sommerfeld-Center for Theoretical Physics and Center for NanoScience, Department of Physics, Ludwig-Maximilians-Universit\"at M\"unchen, Theresienstr. 37, 80333 Munich, Germany; 
$^2$Nanosystems Initiative Munich (NIM), Ludwig-Maximilians-Universit\"{a}t M\"{u}nchen, Schellingstra{\ss}e 4, D-80333 Munich, Germany; $^3$Department of Physics, University of California, San Diego, California 92093, USA}

\begin{abstract}
Here we study a driven lattice gas model for microtubule depolymerizing molecular motors, where traffic jams of motors induce stochastic switching between microtubule growth and shrinkage.
We term this phenomenon \enquote{traffic dynamic instability} because it is reminiscent of microtubule dynamic instability [T. Mitchison and M. Kirschner, Nature 312, 237 (1984)].
The intermittent dynamics of growth and shrinking emerges from the interplay between the arrival of motors at the microtubule tip, motor induced depolymerization, and motor detachment from the tip. The switching dynamics correlates with low and high motor density on the lattice. This leads to an effectively bistable particle density in the system. 
A refined domain wall theory predicts this transient appearance of different phases in the system. The theoretical results are supported by stochastic simulations.
\end{abstract}

\maketitle

\section{Introduction}
Microtubule (MT) depolymerizing enzymes~\cite{Desai1999,Wordeman2005,Walczak2013} are considered important for MT length-regulation~\cite{Howard2007,Melbinger2012,Reese2014}. These enzymes function in parallel to MT dynamic instability~\cite{Desai1997}, which is the hydrolysis-driven stochastic switching of MTs between a growing and a shrinking state~\cite{Mitchison1984,Hill1984,*Hill1984a}. 
The class of MT depolymerizing enzymes~\cite{Walczak2013} contains the kinesin-8 protein family, which are molecular motors that walk towards the MT plus-end~\cite{Varga2006,Gupta2006,Mayr2007,*Stumpff2008,*Tischer2009}. These have been studied in detail in the biological literature, see Ref.~\cite{Su2012} and references therein.

Suitable theoretical models to describe the collective movement of molecular motors on filaments are driven lattice gases~\cite{Derrida1998,Schuetz2001,Blythe2007,Krapivsky2010,Chou2011}. Such models explain the formation of traffic jams on MTs~ \cite{Parmeggiani2003,*Parmeggiani2004,Lipowsky2001,*Klumpp2003,Leduc2012} as observed in experiments~\cite{Leduc2012,Subramanian2013}.
Recently also lattice gases of dynamic system size were studied. For example when individual particles trigger lattice growth by forming new lattice sites at the lattice end~\cite{Sugden2007b,Sugden2007a,Schmitt2011,Muhuri2013}, or remove lattice sites~\cite{Klein2005,Hough2009,Reese2011}.

The interplay between lattice growth and shrinking was investigated in a variety of settings, e.g. when particles stabilize shrinking lattices~\cite{Nowak2007}, or depolymerize growing lattices~\cite{Melbinger2012,Reese2014,Johann2012,*Erlenkaemper2012,Arita2015a}.
An analogy to these molecular motor systems can be found in queuing theory, where the length of a queue is dynamic~\cite{Arita2011,*Arita2012,*Arita2013} and waiting times are of central interest~\cite{Gier2014}.

Here we study a simplified stochastic model for a dynamic MT tip and MT depolymerizing molecular motors. The model is based on the \emph{totally asymmetric simple exclusion process} (TASEP)~\cite{MacDonald1968} and includes specific stochastic processes that account for the growth and shrinking of the lattice.
Particularly, we study a kinetic model where the following MT tip related processes are included: 1) motors depolymerize the MT at the tip, 2) motors inhibit MT growth if bound to the tip, and 3) motors have a finite dwell time at the tip. 
Remarkably, the system investigated here displays stochastic switching between extended periods of MT growth and MT shrinking. This intermittent dynamics is reminiscent of MT dynamic instability~\cite{Mitchison1984}, yet its origin is different and relies on the \emph{dynamics of motors} at the MT tip:
We identify the spontaneous formation of motor traffic jams at the MT tip as a \enquote{catastrophe} event, because it leads to shrinking of the MT. On the other hand, we identify the stochastic detachment of a motor from the tip as a \enquote{rescue} event, because it initiates MT growth in the model.
The predictions of the model can be tested in biochemical reconstitution experiments~\cite{Gell2010} and could help to identify detailed MT-motor interactions as they become increasingly accessible to experiments, see e.g.~Ref.~\cite{MKGardner2011}.

This work is organized as follows. In section~\ref{sec:model} the details of the model are presented.
In the results section we present the phenomenon of \enquote{traffic dynamic instability} (\ref{sec:dynam}). Then, the mean field solution of the model is introduced and discussed (\ref{sec:limit}).
In sections \ref{sec:shock} and \ref{sec:dw} we numerically study the formation of shocks and develop a domain wall theory which quantifies our observations. In particular, this domain wall theory allows to identify metastable regimes in the phase space of the system as is shown in the following sections:
In section \ref{sec:stripe} we identify a phase of stripe formation due to particular velocities of domain wall motion in the system.
The intermittent phases are analyzed in greater detail analytically as well as numerically in sections \ref{sec:interm} and \ref{sec:bistab}. Finally we discuss our results and conclude in section~\ref{sec:discussion}.
\begin{figure}[h]
\centering
\includegraphics[]{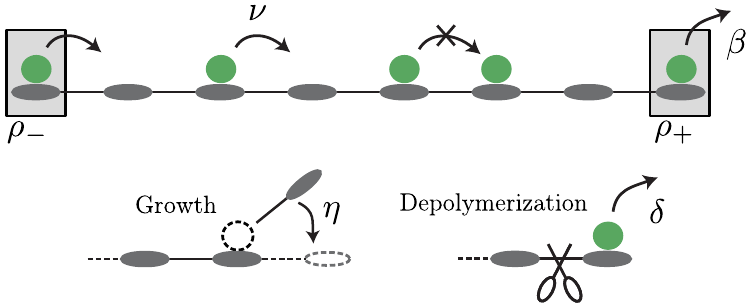}\caption{Molecular motors on the MT lattice. At the left motors enter the lattice from a constant reservoir $\rho_-$, which mimics the motor density in bulk of the MT. The motor density at the tip (MT plus-end) is denoted $\rho_+$. Motors detach from the MT tip at rate $\beta$. The MT grows in \emph{absence} of a motor at rate $\eta$ and is depolymerized by a motor at rate $\delta$. 
\labfig{cartoon}\labfig{model}}
\end{figure}
\section{Model}\label{sec:model}
We consider a lattice gas model for molecular motors close to the MT plus-end as illustrated in Fig.~\reffig{cartoon}. Motors move from the bulk of the MT (left) to the plus-end (right) at rate $\nu=1$ if the next site is empty. This choice of $\nu$ sets the timescale of all other rate constants. The dynamics can be formulated in terms of occupation numbers $n_i$, where $n_i=1$ and $n_i=0$ denote the presence and absence of a particle, respectively. 
At the left, the system is coupled to a constant particle reservoir $\rho_-$, which emulates the bulk of the MT. At the right, the model corresponds to the MT plus-end, where particles detach at rate $\beta$. The above model is known as the \emph{totally asymmetric simple exclusion process} (TASEP) and plays a paradigmatic role in non-equilibrium statistical mechanics~\cite{Krug1991,Derrida1992,Derrida1993,Schuetz1993}. 
Furthermore, we denote the probability that a motor occupies the tip-site with $\rho_+$ and the probability to find a motor in bulk of the lattice $\rho_\tn{b}$. 
We consider a lattice of constant size $N$, which is co-moving with the plus-end of the MT. At the left boundary, the system looses a particle if the lattice grows and the leftmost site is occupied.
In case of a depolymerization event, the lattice moves to the left, where one lattice site is added. The newly added site is occupied with probability $\rho_-$.
At the right boundary, two more processes can happen: \emph{(i)} Particles remove the terminal site with rate $\delta$~\cite{Reese2011}. \emph{(ii)} If the tip-site is \emph{empty} the lattice grows at rate $\eta$~\cite{Reese2014}.
These dependencies of MT growth and shrinking on motor occupation at the tip can also be interpreted as the two different nucleotide states of MT tips~\cite{Mitchison1984}: in the GTP state the filament grows, corresponding to the absence of a motor in our model; in the GDP state the filament shrinks, corresponding to a motor bound to the tip in our model. 
The dynamics of motors on the lattice, however, renders our model different from GTP hydrolysis dependent switching, and, therefore, distinct from MT dynamic instability.
In the model presented here switching between growth and shrinking (catastrophe) is due to the spontaneous formation of traffic jams on growing MTs. Switching from shrinking to growth (rescue) is due to detachment of a motor from the tip.
\begin{figure}
\centering
\includegraphics{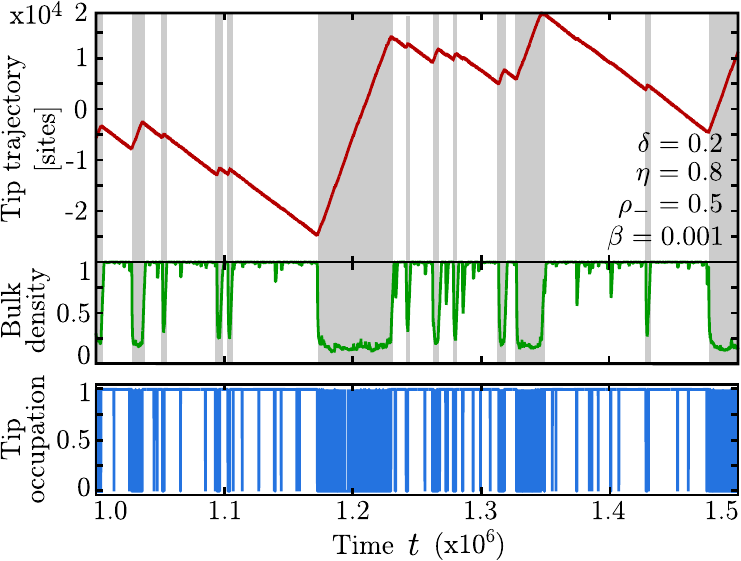}
\caption{Stochastic simulations show intermittent dynamics. Switching between periods of growth and shrinking are reminiscent of MT dynamic instability and indicate bistable behavior of the system: The three panels show the trajectory of the lattice tip, the bulk motor density on the MT $\rho_\tn{b}$, and the motor occupation of the tip $n_+$. Shaded areas indicate periods of growth that involve low bulk density $\rho_\tn{b}\approx0.2$. System size is $N=200$.\labfig{dyna}}
\end{figure}
\section{Results}
\subsection{Intermittent dynamics}\label{sec:dynam}
In stochastic simulations of the model we observe intermittent dynamics. In Fig.~\reffig{dyna} this is illustrated for three key observables, which are the trajectory of the MT tip, the average bulk density of motors on the lattice $\rho_\tn{b}=\tfrac{1}{N}\sum_i^N n_i$, and the motor occupation at the tip $n_+\in\{0;1\}$, see Fig.~\reffig{dyna}.
Periods of depolymerization are distinguished from periods of growth by a high bulk density $\rho_\tn{b}\approx 1$ and persistent occupation of the tip $n_+=1$, cf. the white and shaded areas in Fig.~\reffig{dyna}, respectively. 
The underlying switching processes of the system can be understood intuitively in terms of a separation of timescales: If the tip site is occupied, $n_+=1$, it constitutes a bottleneck~\cite{Pierobon2006,Turci2013} behind which particles pile up due to motor traffic and slow depolymerization. 
For rare tip detachment ($\beta\ll 1$) this bottleneck persists and eventually induces a high density of motors in bulk, $\rho_\tn{b}\approx 1$.
The high bulk density manifests the shrinking state of the filament. 
However, the system may stochastically switch to growth. 
This can happen on the occasion that the bottleneck at the tip site is removed through tip detachment (rate $\beta$).
Then the tip site becomes empty, $n_+=0$, which entails that the lattice is in its growing state. The timescale of this rescue event is $\beta^{-1}$.\\
Similarly, it is possible to obtain a qualitative understanding of catastrophe events:
At high growth rates $\eta$, motors can not keep up with the growing tip. The system assumes a steady state where the bulk density is significantly smaller than in the depolymerizing state, $\rho_\tn{b}\approx 0.2$, see Fig.~\reffig{dyna}. 
Stochastic switching into the shrinking state may happen if a single motor reaches the tip site and induces traffic jam formation, which reverses growth to shrinking.
Thus, the timescale of catastrophe is given by the arrival rate of motors at the tip, $j_+$. This arrival rate is sensitive to current fluctuations and the system size $N$~\cite{Gorissen2012,*Gorissen2011,*Lazarescu2011}.
\begin{figure}
\centering
\includegraphics{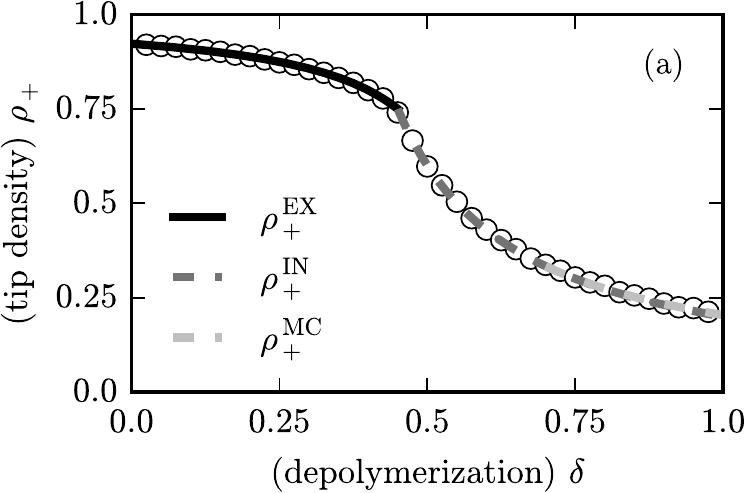}
\includegraphics{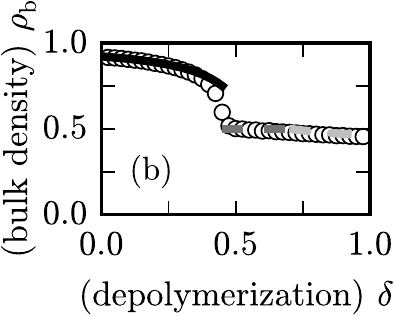}$\,\,\,\,$
\includegraphics{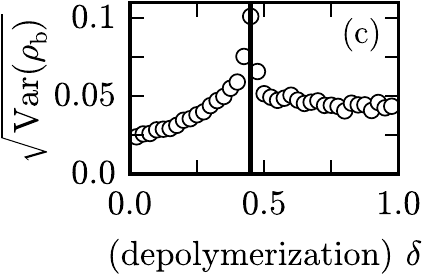}
\caption{(a) Comparison of stochastic simulations (symbols) and analytic results (lines) obtained for the tip densities from mean-field calculations; see Eqs.~\eqref{eq:tipIN}, \eqref{eq:tipEX}, and \eqref{eq:tipMC}.
For $\eta=0.35$, $\rho_-=0.5$, $\beta=0.05$, and $N=400$ we find very good agreement between theory and simulations. The transition between the IN and the MC phase can hardly be recognized; we plotted both functions and elaborate on the exact criterion later. \label{fig:mean-field}
(b) Data for bulk densities (symbols) also confirms the validity of mean-field calculations [lines as in panel (a)]. The transition between the EX and the IN phase is discontinuous in the bulk densities in agreement with what is known from TASEP: At the transition between the EX and the IN phase, both phases coexist and are separated by a diffusing domain-wall (DW). (c) The DW ensues density fluctuations, which we measured in terms of the normalized standard deviation of the bulk density. At the EX/IN-transition density fluctuations show a pronounced peak (indicated by the vertical line). 
\labfig{INEX}}
\end{figure}
\subsection{Limitations of the mean-field approach}\label{sec:limit}
In a first step to understand the dynamics of the system, we determine the tip-densities, $\rho_+$, within a mean-field approach~\cite{Melbinger2012}.
There are three generic phases of motor dynamics on the lattice: a low density phase (IN phase), a high density phase (EX phase), and a maximal current phase (MC phase). 
The dynamics of such a system is likewise dependent on particle input $\rho_-$, the particle exit rate $\beta$, or the capacity of particle flow on the lattice, respectively~\cite{Krug1991,Derrida1992, Derrida1993,Schuetz1993}. 
As shown recently, this requires an analysis of bulk and boundary currents in the system~\cite{Melbinger2012}. 
Employing a mean-field approximation for nearest neighbor occupation numbers, $\avg{n_i n_{i+1}}=\avg{n_i}\avg{ n_{i+1}}$, the bulk current of the system reads~\cite{Reese2014}
\bequ
J_\tn{b}(\rho_\tn{b},\rho_+)=\rho_\tn{b} (1-\rho_\tn{b} ) +\delta  \rho_\tn{b}  \rho_+ -\eta  \rho_\tn{b}  (1-\rho_+),
\label{eq:current}
\eequ 
where the terms on the right hand side stand for particle hopping with on-site exclusion, depolymerization, and polymerization of the MT.
Note that the latter depend on the probability that a motor is bound to the tip.
In terms of particle movements, depolymerization and polymerization correspond to parallel updates of all motors on the lattice towards, or away from the lattice tip, respectively. 

The current of particles that leave the system at the MT tip depends on depolymerization and detachment events and the tip density, 
\bequ
J_\tn{exit}(\rho_+)=(\delta+\beta)\rho_+\, .
\eequ
The tip densities in the IN and the EX phase are obtained in a straightforward manner~\cite{Melbinger2012,Reese2014}. In the IN phase $\rho_\tn{b}=\rho_-$, and in the EX phase $\rho_\tn{b}=\rho_+$. Bulk and tip currents balance in the steady state due to particle conservation, 
\bequ
J_\tn{b}(\rho_\tn{b},\rho_+)=J_\tn{exit}(\rho_+)\, ,
\eequ
and can be solved for the tip density. As results one obtains 
\bequ
\rho_{+}^\tn{IN}=\frac{\rho (\rho +\eta -1)}{\rho  (\delta +\eta )-\delta -\beta }\, ,\label{eq:tipIN}
\eequ
and
\bequ
 \rho_{+}^\tn{EX}=1-\frac{\beta}{1-\delta -\eta } \, .\label{eq:tipEX}
\eequ
\begin{figure*}
\centering
\includegraphics{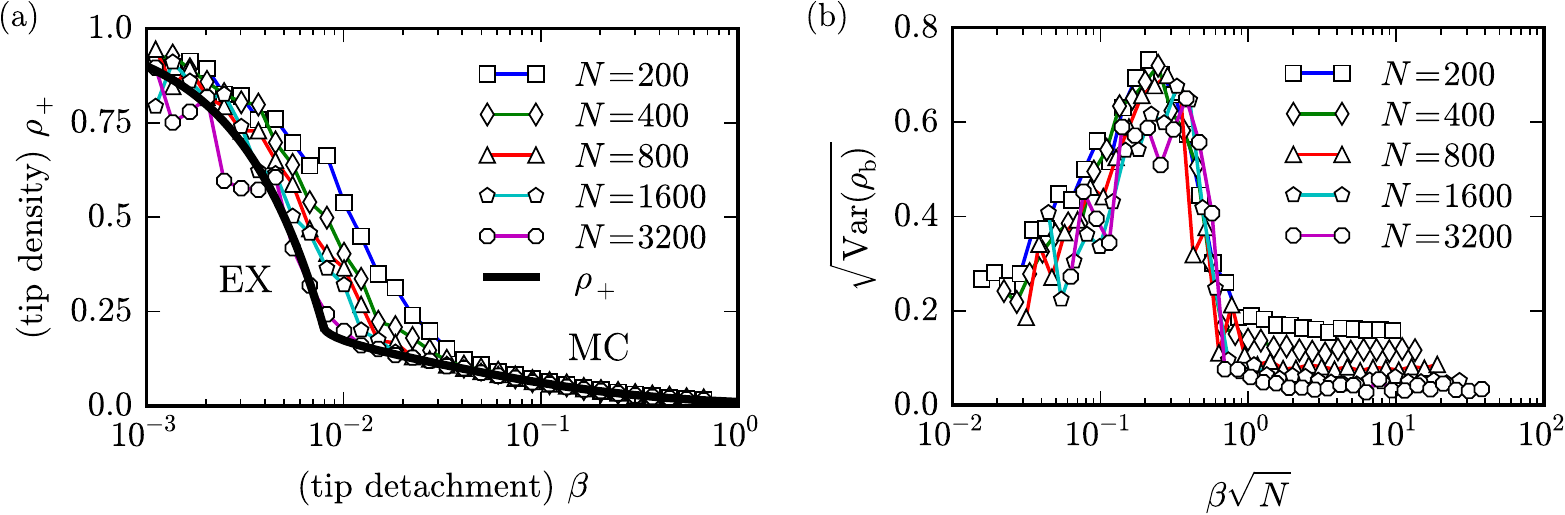}
\includegraphics{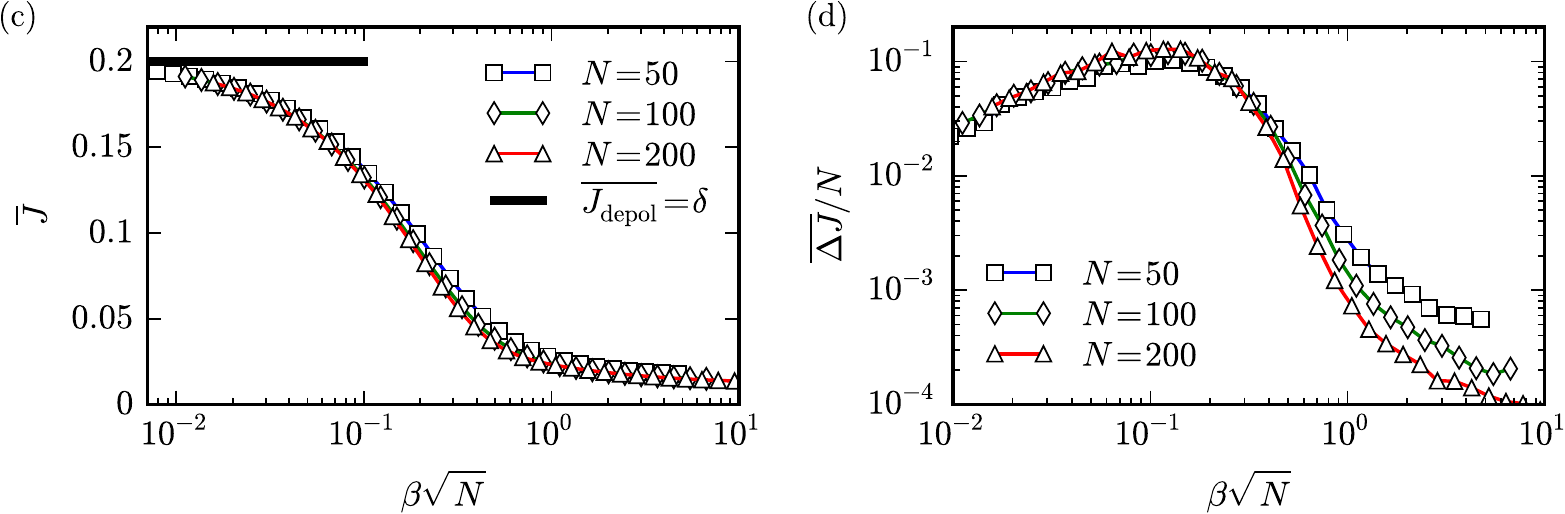}
\caption{Deviation from the mean-field approximation depending on detachment rates $\beta$, for growth rate $\eta=0.79$, and depolymerization rate $\delta=0.2$. Panel (a) shows how the data for tip densities deviates from the analytic results for a set of system sizes. The deviation from mean-field occurs at the transition between the EX phase and the MC phase, Eqs.~\eqref{eq:tipEX} and~\eqref{eq:tipMC}, respectively.  
Note that the deviation from mean-field result depends on the system size $N$.
(b) Bulk density fluctuations in terms of the normalized standard deviation of $\rho_\tn{b}$. An heuristic scaling analysis, where we rescaled the detachment rate as~$\beta\sqrt{N}$, shows data collapse for the onset of characteristic density fluctuations.
(c) The average current through the system $\overline{J}$ measured from stochastic simulations was evaluated from $10^3$ realization of the process, where $\overline{J}=\avg{Q}/\Delta t$. The number of particles that leave the system from the tip through depolymerization or detachment is denoted $Q$ and was measured in time intervals $\Delta t=10^6$.
Scaling is obtained by plotting the data versus~$\beta\sqrt{N}$. Panel (d) shows the scaled current fluctuations $\overline{\Delta J} / N$ of the dataset presented in (c), with $\overline{\Delta J}=(\avg{Q^2}-\avg{Q}^2)/\Delta t $.
The data also reveals scaling with $\beta\sqrt{N}$ and supports the scaling relation between tip detachment and system size.
\label{fig:betaN}}
\end{figure*}
In the MC phase the current through the system is determined by the transport capacity of the lattice defined from an extremal current principle~\cite{Melbinger2012,Reese2014}, $\partial_{\rho_\tn{b}} J_\tn{b}=0$. This condition ensues bulk and tip densities in the MC phase
\begin{widetext}
\bequa
\rho_\tn{b}^\tn{MC}&=&\frac{\beta +\delta -\sqrt{(\beta +\delta ) (\beta +\eta  (\delta +\eta -1))}}{\delta +\eta }\, ,\\
\rho_{+}^\tn{MC}&=&\frac{2 \beta +\delta  (\eta +1)+(\eta -1) \eta -2 \sqrt{(\beta +\delta ) [\beta +\eta  (\delta +\eta -1)]}}{(\delta +\eta )^2} \, . \label{eq:tipMC}
\eequa 
\end{widetext}
To test the validity of the mean-field approach we compare the analytic results for the tip density, $\rho_+$, with stochastic simulation data. 
For slow growth rates we find excellent agreement between the calculations and the data, cf. Fig.~\reffig{mean-field}(a).
The transition between EX and IN phase is discontinuous as known for TASEP, see Fig.~\reffig{mean-field}(b).
For a later comparison with the intermittent regime, we also evaluated characteristic density fluctuations across the EX/IN transition, see Fig.~\reffig{INEX}(c).
These observations are in agreement with the low-density high-density coexistence  in the classical TASEP~\cite{Derrida1992,Derrida1993,Schuetz1993}, which corresponds to the case of $\delta=\eta=0$ and $\rho_-=\beta$ in our model. 
In Fig.~\reffig{INEX} density fluctuations show a pronounced peak at a critical depolymerization rate $\delta_c=\rho_--\beta$, which can be attributed to the EX/IN transition.
Our observation of intermittent behavior, suggests that the mean-field approximation becomes invalid at particular parameter values. Thereby the detachment rate of motors from the tip, $\beta$, plays a critical role.
To understand the emergence of intermittent behavior, and the eventual break-down of the mean-field approximation, we studied the system as a function of system size $N$ and $\beta$ in stochastic simulations~\cite{Gillespie1976,*Gillespie1977}.
Figure~\reffig{betaN}(a) shows how the data for the tip densities $\rho_+$, deviate from the analytic results for a growth rate $\eta=0.79$ and a depolymerization rate $\delta=0.2$. 
While there is good agreement for relatively large motor detachment rates $\beta>0.03$, analytic results and the data deviate significantly for smaller values of $\beta$.
To exclude the possibility that these deviations from mean-field results are due to finite size effects, we also recorded the bulk density and the particle current through the system as well as the fluctuations of these two quantities.
As already shown above in Fig.~\reffig{dyna}, the bulk density switches between states of high and low density in the intermittent regime. Thus, we hypothesize that density and current fluctuations are characteristic for the intermittent regime.
To test this hypothesis we evaluated the bulk density with respect to fluctuations in terms of the normalized standard deviation $\sqrt{\text{Var}(\rho_\tn{b})}=\sqrt{\avg{\rho_\tn{b}^2}-\avg{\rho_\tn{b}}^2}/\avg{\rho_\tn{b}}$.
The results are shown in Fig.~\reffig{betaN}(b). We found that for all studied system sizes, ranging from $N=200$ to $N=3200$, fluctuations peak at a particular value of $\beta$. This excludes the possibility that the observed phenomenon is a finite size effect. Furthermore, by plotting the data versus the rescaled detachment rate, $\beta\sqrt{N}$, we found data collapse for the onset as well as for the peak of the fluctuations. This heuristic analysis ensues the following system size dependent law for the onset of intermittent dynamics
\bequ
\beta_c\propto N^{-1/2}\, .\label{eq:critbeta}
\eequ
The numerical measurement of the average current in the system, $\overline{J}$, and the current fluctuations $\overline{\Delta J}$, confirms the above scaling behavior, see Fig.~\ref{fig:betaN}(c) and (d). 
Note, that the onset of intermittency also depends on the actual values of $\delta$ and $\eta$, as will be shown later.

\subsection{Formation of shocks}\label{sec:shock}
The appearance and dissolution of motor traffic jams at MT tips, provides a mechanism for stochastic switching between states of growth and shrinking. 
Recent studies have investigated TASEP systems with similarly complex shock dynamics, see~\cite{Turci2013,Pinkoviezky2014,Sahoo2015,VandenBerg2015}. 
In the following we briefly discuss methods and results of some of those references and briefly elaborate on the differences to our work. 

\citeauthor{Turci2013}~\cite{Turci2013} investigated a system with a defect which underlies on/off kinetics on an otherwise static lattice with TASEP dynamics. Intermittent density fluctuations are observed, including strong deviations from the classic mean-field approach. 
To improve beyond a mean-field approach, the authors employ an intermittent mean-field approach which allows to calculate the average current in the system during intermittency. 
\citeauthor{Sahoo2015}~\cite{Sahoo2015} investigated a defect which in addition to attachment/detachment kinetics also diffuses. In their model traffic jams form and dissolve stochastically in the bulk of the lattice. The bulk defects considered in the above references~\cite{Turci2013,Sahoo2015} lead to phenomena which are similar to our observations at the MT tip. The difference to our work is that in \cite{Turci2013} and \cite{Sahoo2015} the defects are roadblocks, whereas in our model the kinetic processes at the MT tip can function as a defect.

\citeauthor{Pinkoviezky2014}~\cite{Pinkoviezky2014} investigated a system with defect particles on a constantly growing lattice, where the defect property is passed from one particle to the next against the direction of transport. This ensues defect propagation and interesting DW dynamics in the bulk of the lattice. A simple mean-field approach fails to describe the system, but a careful analysis of the different spatial domains that evolve allows an analysis of the defect dynamics.
The main differences to our model are, the feedback between lattice dynamics and tip occupation, and that the MT tip site is the only defect in the system we study.

In the following we propose a domain wall theory to explore phase diagrams of TASEP systems where tip dynamics and bulk dynamics are coupled through shortening or growth processes.
The theory reveals dynamical phase transitions in the model and thereby explains why simulation data deviate from the mean-field calculations, cf. Fig.~\ref{fig:betaN}(a). 
On the level of individual trajectories, however, the mean-field results can be verified: The velocities of domain walls can be read out directly from kymographs.
Our approach generalizes previous work on domain wall theory~\cite{Kolomeisky1998,Popkov1999} towards an understanding of driven diffusive systems of interacting particles on dynamically evolving lattices.

\subsection{Domain wall theory}\label{sec:dw}
\begin{figure}
\centering
\includegraphics[]{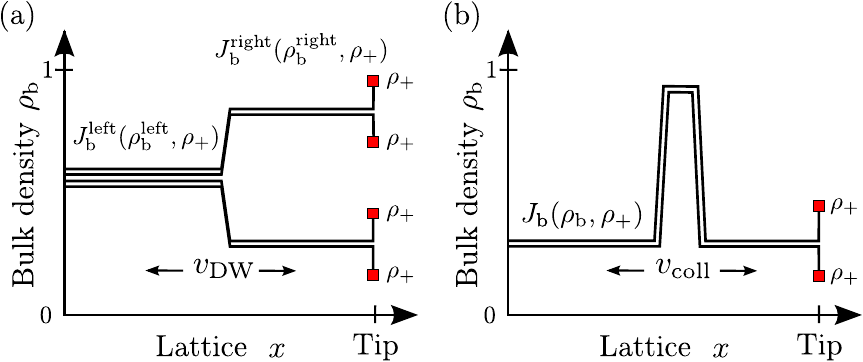}
\caption{Illustration of possible domain walls (a) and a density perturbation (b) in a background density $\rho_\tn{b}$. 
The spike-like behavior of the density profile at the lattice tip is also illustrated. Note the particular role of the tip density $\rho_+$: it influences the currents in the system.
\label{fig:dw}}
\end{figure}

\begin{figure*}
\centering
\includegraphics{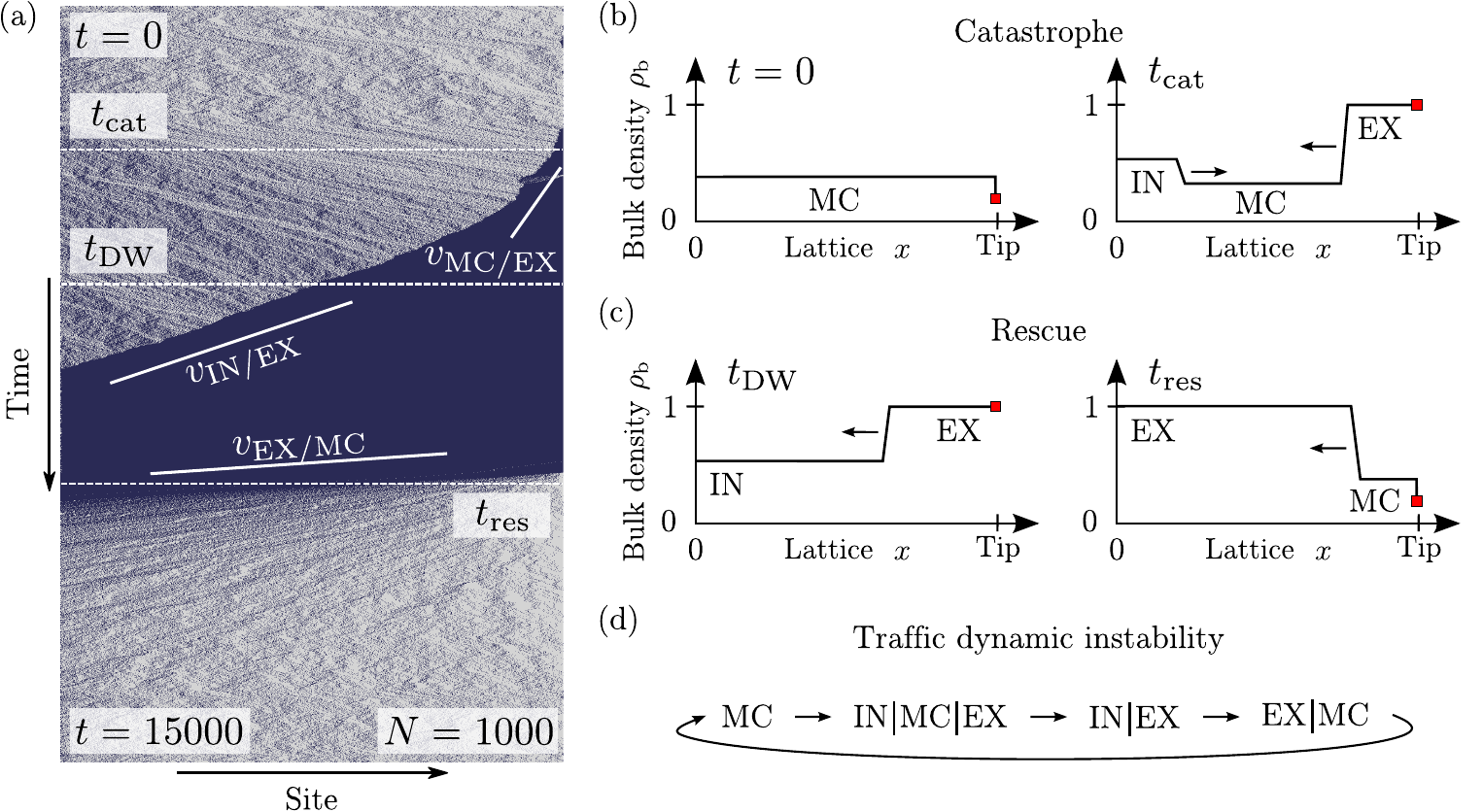}
\caption{(a) Kymograph of stochastic switching from lattice growth to shrinking and back. Solid lines indicate the velocities at which DWs propagate in the bulk of the lattice. For the time points indicated by thin dashed lines the density profiles are illustrated in panels (b) and (c) as indicated.
Panel (b) shows the DW dynamics before and after the catastrophe event. In (c) the DW dynamics before and after the rescue event is illustrated. (d) Illustrates the consecutive phase changes in the system which we call \enquote{traffic dynamic instability}, indicative of the fact that the system does \emph{not} settle into its non-equilibrium stationary states.
Parameters in (a) are $\beta=0.001$, $\rho_-=0.5$, $\delta=0.2$, $\eta=0.79$.
\label{fig:kymo}}
\end{figure*}

Because analytic expressions for the currents, the tip densities, and the bulk densities are known for the different phases, we employ a \emph{domain wall} (DW) theory and \emph{extremal current principle}~\cite{Krug1991,Kolomeisky1998,Popkov1999,Hager2001}. 
In short, DW theory determines how shocks and density perturbations evolve in the system: The direction in which a shock moves and the direction in which a density perturbation spreads determine the existence and stability of the different phases~\cite{Kolomeisky1998}. 
In other words, we are interested in the sign of various DW velocities $v_\tn{DW}$, and the sign the collective velocity $v_\tn{coll}$ which tells about the spreading of density perturbations.
\begin{figure}
\centering
\includegraphics[]{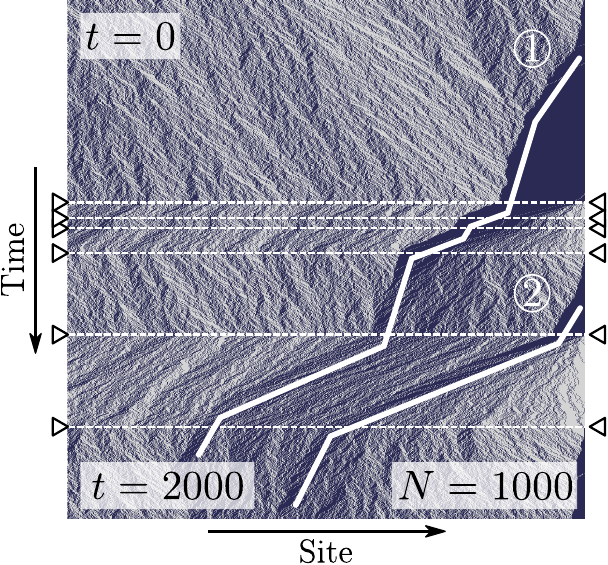}
\caption{Kymograph of the model in a regime with strong density fluctuations. The data shows how two traffic jams nucleate at the MT tip and propagate into the bulk of the system before they dissolve. Dashed lines and triangles indicate switching of the tip occupation between $n_+=1$ and $n_+=0$.
Shock formation and propagation in the bulk of the lattice is highlighted by white solid lines.
Parameters are $\beta=0.005$, $\rho_-=0.5$, $\delta=0.2$, $\eta=0.79$.
\label{fig:multidw}}
\end{figure}
In the model considered here the bulk currents are not independent of the tip density, as in the TASEP.
Microscopically this means that the occupation number of the tip $n_+\in\{0,1\}$ influences DW motion and thus the phase behavior of the system, see the kymograph in Fig.~\ref{fig:multidw} for example.
The figure shows stochastic switching in the tip occupation number. Switching between $n_+=0$ and $n_+=1$ is indicated by arrows and dashed horizontal lines. 
The solid lines are guides to the eye indicating the correlation between tip occupation and distinct DW velocities in the bulk of the lattice.
If switching events occur within a relatively short time window, multiple DWs coexist in the bulk of the lattice as indicated by the numbers $\raisebox{.5pt}{\textcircled{\raisebox{-.9pt} {1}}}$ and $\raisebox{.5pt}{\textcircled{\raisebox{-.9pt} {2}}}$ in Fig.~\ref{fig:multidw}.
The DW velocity can be analyzed analytically, it is given by
\bequ
v_\text{DW}=\frac{J^{\text{left}}-J^{\text{right}}}{\rho^{\text{left}}-\rho^{\text{right}}}\, ,
\eequ
where \emph{left} and \emph{right} denote the densities and currents on either side of a DW. The sign of $v_\text{DW}$  determines whether a shock in the system travels to the left ($v_\text{DW}<0$) or the right ($v_\text{DW}>0$).
For our purposes the typical procedure by~\citeauthor{Kolomeisky1998}~\cite{Kolomeisky1998} needs to be modified, because the currents $J$ on either side of the DW depend explicitly on the tip density $\rho_+$ as illustrated in Fig.~\reffig{dw}.
In the following we investigate $v_\tn{DW}$ between the MC phase and the EX phase, since we have seen that this transition is involved in the stochastic switching between a growing and a shrinking state of the lattice, cf. Fig.~\ref{fig:kymo}.
For the moment, consider the MC phase at the left side of a DW and the EX phase at its right. 
This implies that the tip density is in the EX phase $\rho_+=\rho_+^\text{EX}$. 
The DW velocity $v_\text{left/right}$ then reads 
\bequ
v_\text{MC/EX}=\frac{J(\rho_\tn{b}^\text{MC},\rho_+^\text{EX}) - J(\rho_\tn{b}^\text{EX},\rho_+^\text{EX})}
{\rho_\tn{b}^\text{MC} - \rho_\tn{b}^\text{EX}}\,.
\eequ
Note that both currents, $J^\text{left}$ and $J^\text{right}$, depend on the same tip density, $\rho_+^\text{EX}$, but differ with respect to the bulk densities $\rho^\text{left}=\rho_\tn{b}^\text{MC}$ and $\rho^\text{right}=\rho_\tn{b}^\text{EX}$. 
Vice versa, if the MC phase is on the right side of the DW the tip density is $\rho_+=\rho_+^\text{MC}$. With the EX phase at the left, the DW velocity is 
\bequ
v_\text{EX/MC}=\frac{J(\rho_\tn{b}^\text{EX},\rho_+^\text{MC})-J(\rho_\tn{b}^\text{MC},\rho_+^\text{MC})}
{\rho_\tn{b}^\text{EX}-\rho_\tn{b}^\text{MC}}\,.
\eequ 
Evaluating the DW velocities in the system by employing the current given in Eq.~\eqref{eq:current} and the tip and bulk densities for the individual phases, we are able to construct the phase diagram. In the following we provide an attempt for a characterization of the arising phases.
\begin{figure}
\centering
\includegraphics[]{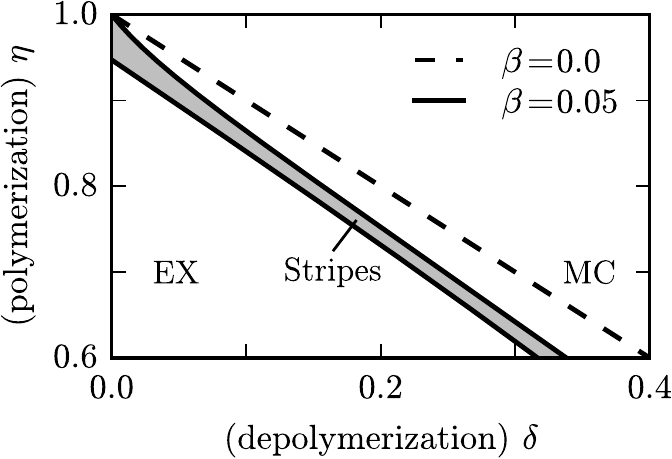}
\caption{Phase diagram for the stripe phase. For $\beta=0$ the system shows a discontinuous transition between EX and MC phase, where stripes emerge right at the critical line (dashed). In the case $\beta>0$ a distinct region in phase space exists in which the system robustly forms stripe patterns due to the creation of shocks at the MT plus end and a subsequent switch to polymerization. 
\label{fig:pd_stripe}}
\end{figure}

\subsection{Stripe phase}\label{sec:stripe}
As suggested from the data shown in the kymograph of Fig.~\ref{fig:kymo}, there is a regime in which the system completely switches from the MC phase to the EX phase for small~$\beta$. 
Because the switching appears as a band of motor particles across the lattice, we refer to this regime as a stripe phase, cf. Fig.~\ref{fig:kymo}(a).
A reasonable and simple condition for the stripe phase is that all three domain walls need a negative DW velocity. This is evident from the kymograph of Fig.~\ref{fig:kymo}.
For $\beta=0$ this condition cannot be fulfilled:
\bequa
v_\tn{IN/EX}&=&-\rho+\delta \, ,\\
v_\tn{EX/MC}&=&-\eta \, ,\label{eq:togrowth}\\
v_\tn{MC/EX}&=&0\, ,\label{eq:todepol}
\eequa
where right after the switching event to growth [Eq.~\eqref{eq:togrowth}] we assumed $\rho_\tn{b}^\tn{right}=\rho_+=0$, and right after the switching event to depolymerization [Eq.~\eqref{eq:todepol}] we assumed $\rho_\tn{b}^\tn{right}=\rho_+=1$, as suggested from the data.
The first two of the above equations indeed show a negative DW velocity as expected. The MC/EX domain wall, however, does not show a negative velocity as observed in the simulations.
Let us note that for $\beta=0$ we find a critical growth rate $\eta^s=1-\delta$ (dashed line in Fig.~\reffig{pd_stripe}) or likewise, a critical depolymerization rate $\delta^s=1-\eta$~\cite{Reese2014}.
To better explore the possibility of $v_\tn{MC/EX}<0$ we study the case $\beta>0$.
We find that for finite values of $\beta$ the critical line broadens up into a regime in which $v_\tn{MC/EX}<0$ is possible.
In terms of the growth rate, the conditions for this regime can be found from $v_\tn{MC/EX}=0$ and that $v_\tn{MC/EX}$ has to be real in the relevant parameter regime. It follows that the parameter region of $v_\tn{MC/EX}\leq 0$ is given by
\bequa
\eta &\geq &\frac 12 \left(1-\delta+\sqrt{-4\beta + (1-\delta)^2} \right) \, ,  \\ 
\eta &\leq & 
1-\delta  \left(1+\frac{\beta }{\beta -\delta ^2+\delta }\right)
 \, .
\eequa
The above relations provide the condition for stripe formation in the system based solely on domain wall theory, see shaded are in Fig.~\reffig{pd_stripe}.
Although we find qualitative agreement with individual kymographs, compare Fig.~\reffig{kymo}(a) and illustrations in Fig.~\reffig{kymo}(b), (c) and (d), the stochastic nature of the phenomenon makes it hard to quantitatively confirm this regime numerically.
The major difficulty thereby is that the regime of intermittent dynamics is not only confined to the stripe phase but extends to above and below in phase space. In these regimes the system does not reach a non-equilibrium steady state either, but is continuously driven between different phases in course of time.
This view is fortified in the following section. 

\subsection{Intermittent phase}\label{sec:interm}
To characterize the non-equilibrium state in which the system does not settle into a particular (non-equilibrium) stationary state, we distinguish domain walls between different phases considering the IN, EX, and MC phase. An illustration of the possible domain walls in the system is shown in Fig.~\ref{fig:dw}(a).
In particular, we are interested in the direction of DW motion. This quantity determines if a transient DW moves to the left or to the right in the system.

Thus, we investigated the sign of $v_\tn{DW}$ for the six possible combinations of DWs in the system.
Our results are summarized in Tab.~\ref{tab:dw}.
In the following we discuss the cases related to the EX phase, because they are particularly relevant to understand the intermittent regime.
We begin with the DW between the EX and the MC phase. 
Notably neither of the two phases, MC or EX at the right hand side of the system, is stable against the other phase on the left hand side of the system. This means, that depending on the initial preparation of the system, MC/EX or EX/MC, the final state of the system is MC or EX, respectively, and thus ergodicity seems to be broken. Similar observations were made recently for an exclusive queuing model~\cite{Schultens2015}. 
The parameter regime where the phenomenon occurs can be determined from the conditions $v_\text{MC/EX}>0$ and $v_\text{EX/MC}>0$. We find a critical growth rate $\eta^c$, which reads
\bequ
\eta^c=1-\delta-\frac{\beta}{1-\delta-\beta}\, . \label{eq:EXMC}
\eequ
Further, the criterion for the EX/MC phase transition, in the absence of particle detachment from the tip, coincides with the change of sign in $v_\mathrm{EX/MC}$ from positive to negative:
\bequ 
\eta^*=1-\delta\, .
\eequ
Consequently for $\eta>\eta^*$ the MC phase is stable as noted previously~\cite{Reese2014}. The system behavior for $\eta<\eta^c$ remains to be determined.
As shown in Tab.~\ref{tab:dw} there is also the possibility that an IN phase exists in the system below a critical growth rate $\eta<\eta^c_\mathrm{IN}$. This finding is in agreement with our earlier observation that density and current fluctuations are large in this parameter regime, cf. Fig.~\ref{fig:betaN}. Here we refrain from a more detailed investigation of the IN phase and study the role of fluctuations for the EX phase instead, because only if the EX phase is \emph{unstable} intermittency is possible.
\begin{table}
\caption{Summary of results for DW motion. $\eta^c$ refers to the value determined by the EX/MC case given by Eq.~\eqref{eq:EXMC}. The boxed cases indicate mutually unstable cases of DW motion.
\label{tab:dw}}
\setlength{\extrarowheight}{5.pt}
\begin{tabular*}{\columnwidth}{@{\extracolsep{\fill}}lccc}
\hline\hline
Domain wall  & $\eta<\eta^c$ & $\eta^c<\eta<\eta^*$& $\eta>\eta^*$ \vspace{.25cm} \\
\hline
$\mathrm{MC}|\mathrm{EX}$ & n.a. & \topinset{$\rightarrow$}{$\textifsym{L|H}$}{3pt}{} & \topinset{$\rightarrow$}{$\textifsym{L|H}$}{3pt}{}   \\

$\mathrm{EX}|\mathrm{MC}$ & n.a. &\topinset{$\rightarrow$}{$\textifsym{H|L}$}{3pt}{} &\topinset{$\leftarrow$}{$\textifsym{H|L}$}{3pt}{} \\
\hline
\vspace{.5cm}
 &  & \framebox{MC \& EX} & MC \\

\hline
$\mathrm{IN}|\mathrm{EX}$ & \topinset{$\leftarrow$}{$\textifsym{L|H}$}{3pt}{} &\topinset{$\leftarrow$}{$\textifsym{L|H}$}{3pt}{} & \topinset{$\leftarrow$}{$\textifsym{L|H}$}{3pt}{} \\

$\mathrm{EX}|\mathrm{IN}$ & \topinset{$\leftarrow$}{$\textifsym{H|L}$}{3pt}{} & \topinset{$\rightarrow$}{$\textifsym{H|L}$}{3pt}{} & \topinset{$\leftarrow$}{$\textifsym{H|L}$}{3pt}{} \\
\hline
\vspace{.5cm}
 &  \framebox{EX \& IN} & EX & {EX \& IN}\footnote{Note this combination is not realistic, because neither the EX nor the IN phase exist in this parameter regime~\cite{Reese2014}.}\\
\hline
$\mathrm{MC}|\mathrm{IN}$ & n.a. & \topinset{$\rightarrow$}{$\textifsym{L|H}$}{3pt}{} &\topinset{$\rightarrow$}{$\textifsym{L|H}$}{3pt}{} \\

$\mathrm{IN}|\mathrm{MC}$ & n.a. & \topinset{$\leftarrow$}{$\textifsym{H|L}$}{3pt}{} &\topinset{$\leftarrow$}{$\textifsym{H|L}$}{3pt}{} \\
\hline
\vspace{.25cm}
 &  & MC & MC \\
\hline\hline
\end{tabular*}
\end{table}
To this end we analyze the collective velocity in the EX phase. The collective velocity is defined as
\bequ
v_\tn{coll}=\partial_{\rho_\tn{b}} J_\tn{b}(\rho_\tn{b},\rho_+)\, . 
\eequ
It probes the stability of a small density perturbation in a background bulk density, see Fig.~\ref{fig:dw}(b) for an illustration.
Quite generally, the EX phase is characterized by negative collective velocity, $v_\tn{coll}^{\tn{EX}}<0$, because in a high background density a small perturbation moves to the left~\cite{Kolomeisky1998}. 
However, in contrast to typical TASEP systems, this is not the case in our model. In the regime of $\eta^c<\eta<\eta^*$, the velocity of a perturbation is positive $v_\tn{coll}^{\tn{EX}}>0$, indicating that density perturbations in the system travel to the right.
Interestingly the critical growth rate which determines $v_\tn{coll}^{\tn{EX}}=0$ coincides with the critical growth rate as obtained from the DW analysis above, $\eta^c$. 

Until now we have relied on the principles of classical DW analysis to determine the phase behavior of the system. 
We have learned that between the EX and the MC phase, for $\eta^c<\eta<\eta^*$, DW velocities and the collective velocity of the EX phase show interesting behavior.
For $\eta>\eta^*$ the system is robustly in the MC phase~\cite{Reese2014}, and for $\eta<\eta^c$ the EX phase is stable in the system.
In the case $\eta^c<\eta<\eta^*$, however, the system does not settle into a stationary state, but is continuously driven between different transient states. 
So far this behavior can be summarized as follows
\bequ
\begin{array}{cc}
\eta > \eta^* \quad\qquad\qquad & \mathrm{MC}\, ,\nonumber \\
\eta^c < \eta < \eta^* \quad\qquad\qquad & \mathrm{MC\, \& \, EX}\, ,\nonumber \\
\eta < \eta^c \quad\qquad\qquad & \mathrm{EX}\, .\nonumber
\end{array}
\eequ

As discussed in the previous sections, it appears from stochastic realizations of the system (Fig.~\ref{fig:multidw}), that perturbations at the MT tip render the bulk of the system unstable and promote the observed intermittent behavior of the system. This importance of the boundaries in driven diffusive systems was recognized by \citeauthor{Krug1991}~\cite{Krug1991}. And the model we study here adds dynamic complexity to boundary induced phase transitions~\cite{Krug1991}.
Therefore we try to include stochastic switching at the MT tip into the analysis of the collective velocity as discussed above.
In the following we show how the collective velocity is affected from switching events at the tip, by assuming that the tip density $\rho_+$ is characterized by only two states, given by the tip occupation number, $n_+=1$ and $n_+=0$. Thus we assume that the tip density takes only values $\rho_+=1$ and $\rho_+=0$, which leads to the following equations for the collective velocity:
\bequa
v_\tn{coll}(\rho_+=1)=1-2\rho_\mathrm{b} + \delta\, ,\label{eq:vcoll_occ}\\
v_\tn{coll}(\rho_+=0)=1-2\rho_\mathrm{b} - \eta \, .\label{eq:vcoll_empty} 
\eequa
\begin{figure*}
\centering
\includegraphics[]{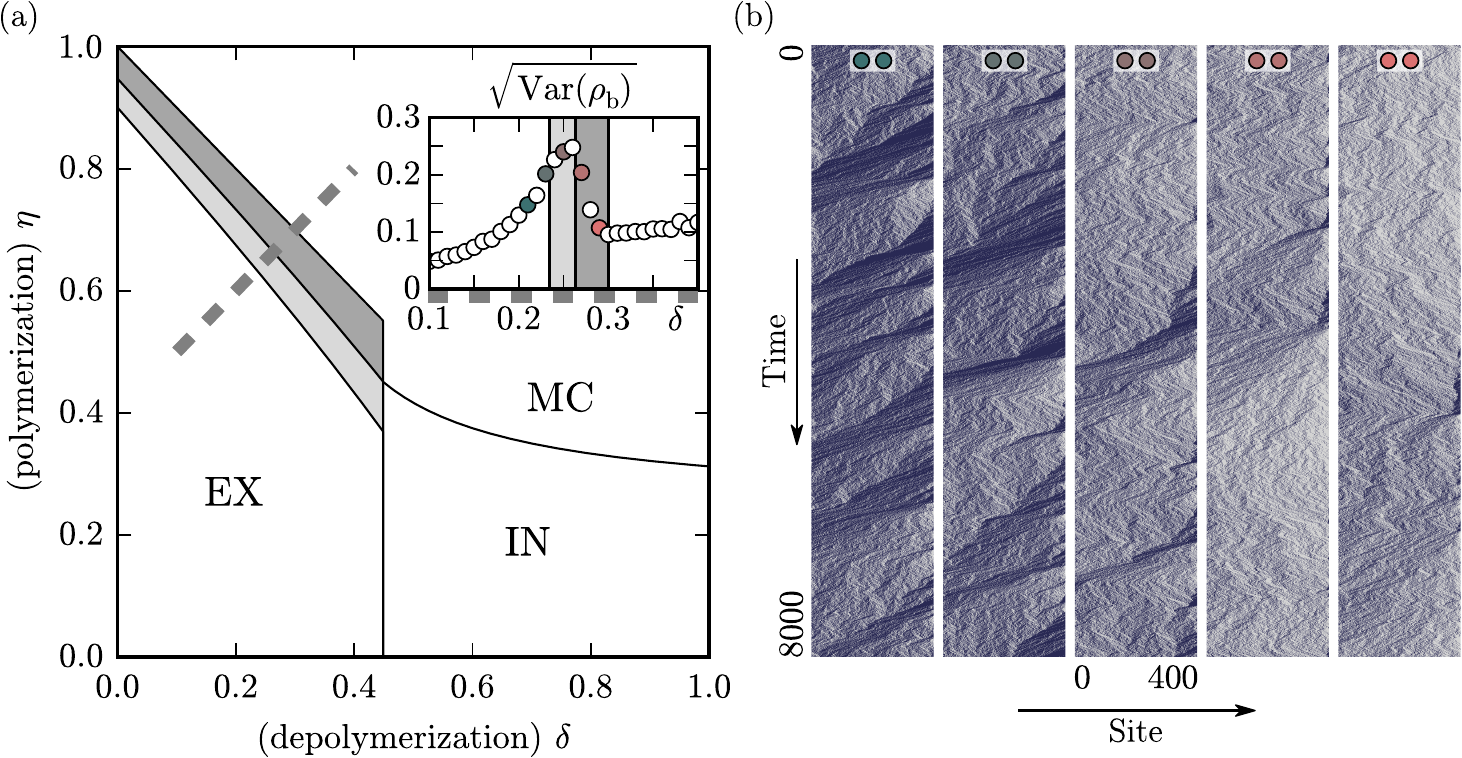}
\caption{(a) Phase diagram of the model in terms of depolymerization and growth rates for $\rho_-=0.5$ and $\beta=0.05$. 
The shaded areas show regimes of intermittent dynamics as calculated in the main text. 
The topology of the diagram with respect to high density (EX), low density (IN) and maximal current density (MC) phases correspond to the case of $\beta=0$~\cite{Reese2014}. 
The inset shows how bulk density fluctuations are increased in the intermittent regime. A comparison with Fig.~\ref{fig:mean-field}(c) reveals that fluctuations are enhanced by almost a factor of four as compared to the discontinuous EX/IN transition.
Data was recorded for system size $N=400$ and $\eta=0.4+\delta$ along the thick dashed line. (b)~Representative kymographs across the intermittent regime. Parameter values correspond to those of the filled circles in the inset of panel (a). \label{fig:var}\label{fig:phases}}
\end{figure*}
The above equations illustrate that stochastic switching between an empty and an occupied MT tip can promptly affect the sign of the collective velocity. This can be seen immediately, because depolymerization (rate $\delta$) and polymerization (rate $\eta$) contribute with different signs to $v_\tn{coll}$.
This means that depending on the state of the MT tip -- occupied or empty -- perturbations may either stabilize or destabilize the bulk density.
This can be illustrated in a straight forward manner for the IN phase, when we choose $\rho_\tn{b}=\rho_-=1/2$ for example. In this case the sign of $v_\tn{coll}$ depends \emph{only} on the tip density:
For $\rho_+=1$ and $\rho_\tn{b}=\rho_\tn{b}^{\mathrm{IN}}$ the lattice is in the shrinking state, while particles on the lattice still travel towards the tip at unit velocity. Consequently perturbation travel to the right with $v_\tn{coll}=\delta$. This follows directly from Eq.~\eqref{eq:vcoll_occ} and is in line with the kymographic data presented above. 
For $\rho_+=0$ the lattice is in the growing state, and particles on the lattice travel towards the tip at a reduced velocity $1-\eta$. As can be seen from Eq.~\eqref{eq:vcoll_empty}, density perturbation travel to the left with negative collective velocity $v_\tn{coll}=-\eta$, again in agreement with the kymographic data. 
Now let us apply this argument for the EX phase, where the situation is more intricate. The sign-change of the collective velocity, $v_\tn{coll}(\rho_\text{b}^\text{EX},\rho_+=0)<0$ and $v_\tn{coll}(\rho_\text{b}^\text{EX},\rho_+=1)>0$, is restricted to a particular regime $\eta^\dagger<\eta<\eta^c$, with 
\bequ
\eta^\dagger=1-\delta-\frac{2\beta}{1-\delta}\, ,
\eequ 
while the remaining part of the parameter space is not affected.
As a consequence the EX phase is destabilized by molecular switching events at the MT tip  in this regime.
This means that for growth rates $\eta^\dagger<\eta<\eta^c$ the molecular noise due to motor occupation at the tip renders the EX phase unstable.
This mechanism is complementary to the above argument for DW velocities. 
In Fig.~\ref{fig:phases}(a) we show the phase diagram with the EX, MC and IN phase as well as the different intermittent regimes. In the inset we quantified density fluctuations across the intermittent region in phase space. The different regimes, $\eta^\dagger<\eta<\eta^c$ (light gray) and $\eta^c<\eta<\eta^*$ (darker gray), can be identified reasonably well, given the complexity of the dynamics. As will be shown in the next section it is instrumental to employ higher order moments and a direct evaluation of bulk density distributions to distinguish between the regimes.
Figure~\ref{fig:phases}(b) shows kymographic data of typical trajectories in the intermittent regime. This direct visualization of the process is hard to analyze computationally, but in the eye of the beholder, domain walls and qualitative differences in current fluctuations can be readily recognized.
\begin{figure}
\centering
\includegraphics{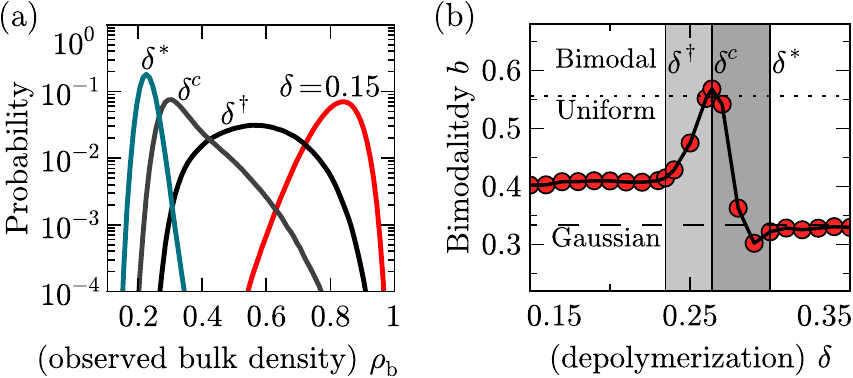}
\caption{(a) Distribution of the bulk density from trajectories as long as $10^8$ time steps for $\beta=0.05$ (thick). (b) The bimodality parameter $b$, supports the role of fluctuations during traffic dynamic instability: $b=1/3$ corresponds to a Gaussian distribution and $b=5/9$ to a uniform distribution. A distribution with $b>5/9$, as reached for $\delta^c$, can be considered bimodal. 
Simulations were conducted at system size $N=400$ and $\eta=0.4+\delta$.\labfig{hist}\labfig{bimodality}}
\end{figure}

\subsection{Bistability and collective motion}\label{sec:bistab}
To complete the picture of the intermittent regime, we performed extensive stochastic simulations to obtain the bulk density distribution, see Fig.~\reffig{hist}(a).
We further analyze the data using the bimodality parameter $b=(\mu_3^2 + 1)/\mu_4$, where $\mu_{3,4}$ are the third and fourth standardized central moments of the density distribution.
Figure~\reffig{hist}(b) displays the results. The intermittent regime is indicate by light and dark gray regions in the plot.
At $\delta^\dagger$, $b$ increases with increasing $\delta$ and $\eta$, it peaks at $\delta^c$ and drops to a constant value $b=1/3$ for $\delta >\delta^*$. 
The latter value corresponds to a Gaussian distribution.\\
The above analysis corroborates our findings for the intermittent regime: 
Stochastic switching between different phases ensues large density fluctuations which can be recognized in terms of a bistable bulk density distribution. 

\begin{figure*}
\centering
\includegraphics[]{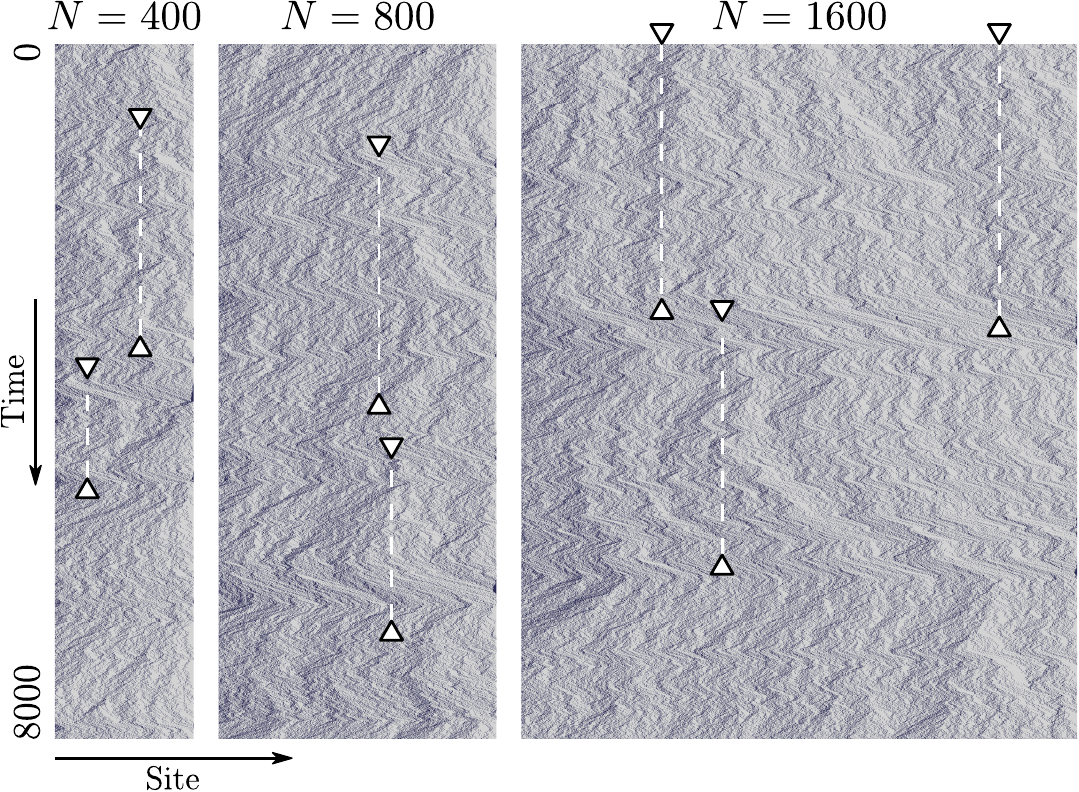}
\caption{At the system size dependent critical detachment rate $\beta_c(N)$, perturbations underlie erratic zig-zag motion in the bulk of the system as highlighted by the triangles and vertical dashed lines. Shown are kymographic data for several system sizes as indicated. Periods of depolymerization and polymerization alternate in a way that density perturbations are maintained within the system. Perturbations seem to perform random walks that are created and annihilated stochastically. Simulation parameters were $\delta=0.2$, $\eta=0.79$, $\rho_-=0.5$, and, from left to right, $\beta_c(400)=0.0125$, $\beta_c(800)=0.008838$, $\beta_c(1600)=0.00625$. 
\label{fig:morekymos}}
\end{figure*}

Finally, we wish to address the scaling relation we have found for density and current fluctuations in the beginning of the paper [Fig.~\ref{fig:betaN}]. 
Although we have discussed the origin of fluctuations and intermittent dynamics in the system, an explanation for the characteristic scaling $\beta_c \propto {N}^{-1/2}$ for the onset of density fluctuations has remained elusive.
In order to gain at least some phenomenological insight, kymographic data was recorded for a set of system sizes and $\beta=\frac 14  {N}^{-1/2}$, what approximately corresponds to the onset of the regime of strong fluctuation.
The results are shown in Fig.~\ref{fig:morekymos}. In particular, we chose the fields of view in a mainly growing state, where the system is most likely in the MC phase. A close inspection reveals that density fluctuations on the lattice do not vanish in a diffusive manner, but rather perform erratic zig-zag motion on the lattice, as indicated by the symbols. This is opposed to the situation $\beta>\beta_c$, where perturbations drift to the left of the system. Similarly for $\beta<\beta_c$ the shrinking phase becomes more likely and perturbations move towards the tip of the lattice.
The data shown in Fig.~\ref{fig:morekymos} suggests that individual density fluctuations are maintained and held in the bulk of the system through the combined dynamics of growth and shrinking.
A detailed analysis however lies beyond the scope of this paper. We think that a theory in which perturbations were considered as quasi-particles or collective excitations of the system are likely to provide a deeper understanding of the $N^{-1/2}$ relation.   

\section{Discussion and Conclusion}\label{sec:discussion}
In this article we presented a driven lattice gas inspired from microtubule depolymerizing motors (kinesin-8), and based on the \emph{totally asymmetric simple exclusion process}. The system exhibits boundary-induced intermittent dynamics, with stochastic switching between different phases of motor traffic and between lattice growth and shrinking. 
The standard mean-field approach explains density and current profiles of the system, but misses a proper description of the intermittent regime. We introduce an extended domain wall theory complemented by an extremal current principle which predicts the intermittent regime, in which multiple phases coexist. 
This phase coexistence can not be resolved with time- or ensemble-averages, because in such data, it appears the system deviates from the usually well-behaved mean-field case.
However, an interpretation of individual realizations of the process (like in a single molecule experiment) allows to apply the mean-field results to the data.

Intermittency in driven systems has been reported before. For example in bidirectional transport~\cite{Evans1995,Willmann2005,*Grossinsky2007,Jelic2012}, and in cellular automata for traffic flow \cite{Barlovic1998,Appert2001}. The reasons for the interesting dynamics in these systems are particle interactions in bulk of the systems -- in contrast to the situation considered here, where the system is triggered at the boundary. 
Quite generally, the presence of bottlenecks leads to interesting effects which affect the particle densities in driven systems. This includes bottlenecks in bulk~\cite{Turci2013,Jelic2012}, at system boundaries~\cite{Popkov2008,Wood2009,Ito2014}, as well as dynamically evolving bottlenecks~\cite{Pinkoviezky2014,Sahoo2015,VandenBerg2015}.
In our case we could attribute the intermittent dynamics to bulk density fluctuations which arise through molecular noise at the lattice tip, i.e. switching between tip occupations $n_+=1$ and $n_+=0$. The sources of this noise is motor detachment and arrival at the tip.
In our model, the bulk density fluctuations can be attributed to a region in phase space in which different phases of motor traffic coexist, or in other words, multiple phases can be transiently stable for a given set of parameters.
To identify these dynamical phase transitions and track them analytically we neglected attachment and detachment kinetics of the motors, and thus assumed a constant density profile. 

In a biological situation where molecular motors interact with MTs, this assumption is valid when the length of the filament exceeds a critical length $\propto {\omega_\tn{a}}^{-1}$, where $\omega_\tn{a}$ is the attachment rate of motors to the MT~\cite{Reese2011}. We checked numerically that the intermittent regime also occurs in the case of a constant density profile with explicit motor attachment and detachment.
This can be understood from a lattice gas point of view, where motor detachment from the tip is equivalent to the creation of a \enquote{hole} at the tip: A hole can reach the tip by particle detachment in bulk and subsequent transport of the hole to the tip through depolymerization of the lattice.

The stochastic switching between growth and shrinking in our model emerges from collective effects of molecular motor traffic.
For MT dynamic instability in contrast the mechanism of stochastic switching between growing and shrinking can be attributed to the nucleotide state of the MT lattice~\cite{Gardner2013}. 
A comparison by numbers could be thought of within the mathematical framework provided by \citeauthor{Dogterom1993}~\cite{Dogterom1993}, where the parameters are growth and shrinking speeds $v_+$, $v_-$, and the frequencies of catastrophe $f_{+-}$ and rescue $f_{+-}$.
Our model relates microscopically to these macroscopic parameters. The speed of the MT tip is a function of the motor density at the MT tip and given by $v(\rho_+)=\eta(1-\rho_+) - \delta \rho_+$~\cite{Reese2014}. Depending on whether the system is in a high density phase or a low density phase $v(\rho_+)$ can be negative or positive. For the switching frequencies the mapping is only possible for the rescue frequency. It can be directly attributed to the parameter of particle detachment from the tip $f_{-+}\propto\beta$. Catastrophe in contrast is initiated by particle arrival at the tip and initiation of a traffic jam. Thereby current fluctuations are important and more elaborate techniques are necessary to calculate these~\cite{Gorissen2012,*Gorissen2011,*Lazarescu2011}. 

A future challenge is to understand the interplay between the MT lattice, MT tips, and MT associated proteins. The characteristics of each of these parts contribute significantly to MT dynamics, but the relations between them are still obscure; necessary information is available:
protein localization mechanisms at MT tips~\cite{Reithmann2015}, enzymatic functions of tip related enzymes~\cite{Zanic2013,Duellberg2014}, and dynamic information about the MT tip structure~\cite{Maurer2014,Alushin2014}.

The present model constitutes an example how enzymes may influence MT dynamics. The mechanism we found is complementary to known mechanisms of MT regulation, but may contribute to MT regulation in a similar way as does MT dynamic instability~\cite{Mitchison1984}.

\begin{acknowledgments}
L.R. is grateful to Matthias Rank and Emanuel Reithmann for numerous valuable discussions and input on a previous version of this manuscript and to Anatoly B. Kolomeisky for an inspiring discussion in the prelude of this work and comments on the manuscript. Financial support by the Deutsche Forschungsgemeinschaft in the framework of the SFB 863, project number B02, is gratefully acknowledged.
\end{acknowledgments}

\end{document}